\documentclass[aps,prb,twocolumn,groupedaddress,showpacs]{revtex4}
\usepackage{graphicx}% Include figure files

\begin{document}

\title{Generalized drift-diffusion model for miniband superlattices}

\author{ L. L. Bonilla\cite{bonilla:email},  R. Escobedo  }

\affiliation{Departamento de Matem\'{a}ticas, Escuela Polit\'ecnica
Superior,  Universidad Carlos III de Madrid, Avenida de la Universidad 30,
28911 Legan{\'e}s, Spain}

\author{ \'Alvaro Perales }

\affiliation{Departamento de Autom\'{a}tica, Escuela Polit\'ecnica, 
Universidad de Alcal\'a, 28871 Alcal{\'a} de Henares, Spain}

\date{ \today  }

\begin{abstract}
A drift-diffusion model of miniband transport in strongly coupled superlattices is derived 
from the single-miniband Boltzmann-Poisson transport equation with a BGK 
(Bhatnagar-Gross-Krook) collision term. We use a consistent Chapman-Enskog method 
to analyze the hyperbolic limit, at which collision and electric field terms dominate the 
other terms in the Boltzmann equation. The reduced equation is of the drift-diffusion type, 
but it includes additional terms, and diffusion and drift do not obey the Einstein relation 
except in the limit of high temperatures. 
\end{abstract}


\maketitle

In recent years, nonlinear charge transport in semiconductor superlattices (SLs) has blossomed 
as a field, driven by the availability of many experimental results and by theoretical 
analyses and simulations of rate equation models \cite{gra95,bon02,wac02}. As it often 
happens, these models have not been derived from more fundamental ``first principles 
formulations'' such as kinetic theory. The situation is different depending on whether the 
SLs are weakly or strongly coupled. In weakly coupled SLs, neighboring quantum wells 
are separated by ``thick'' barriers and vertical transport occurs via sequential resonant 
tunneling through them. Provided intersubband scattering is much faster than escape 
times from a quantum well and the latter are much smaller than dielectric relaxation times, 
electrons are at local equilibrium in the subband of lowest energy \cite{bon02}. Then the 
tunneling current density across a barrier under stationary conditions can be calculated from 
``first principles'' using the Transfer Hamiltonian method \cite{THM}, Green functions for 
a SL under a constant external field \cite{wac02}, etc. This tunneling current (that depends 
on the electron density in the two quantum wells separated by the barrier and on the local 
value of the electric field) is then {\em inserted} in a discrete rate equation model including 
charge continuity and a discrete Poisson equation \cite{bon02,wac02}. No derivation of this 
very reasonable model seems to be known to this date, although its validity has been 
corroborated by numerous experiments. 

In strongly coupled SLs, barriers are ``thin'', minibands are wide, and quantum wells cannot 
be considered as separate entities. Practical models to analyze nonlinear transport are of 
the drift-diffusion type \cite{sib90} or hydrodynamic models \cite{but77,lei91}. 
Drift-diffusion equation (DDE) models typically use a drift velocity obtained from a 
simplified kinetic equation and a diffusion coefficient that obeys the Einstein relation 
\cite{sch02}. The resulting model is a variant of the well-known Kroemer DDE for the 
Gunn effect in bulk n-GaAs \cite{kro66}. For the large fields involved and for the 
non-parabolic SL miniband energy, using an Einstein relation to figure out the diffusion 
coefficient is questionable \cite{bry03} and, in fact, incorrect except in a particular limit. 
Hydrodynamic models are considerably more complicated and have been solved numerically,
but not many analyses of them have been carried out. The same applies to quantum diffusion 
theories \cite{bry03}. ``First-principles derivations'' typically solve a kinetic equation 
numerically or approximately assuming a constant applied electric field and ignoring 
space and time dependence. Then a current density across the SL is calculated for different 
values of the field, and a drift velocity and a diffusion coefficient are figured out. These 
functions are then inserted in a DDE. Results that are ``valid for any type of SL'' typically 
mean that stationary, space-independent solutions of a sufficiently general kinetic equation 
have been found numerically \cite{wac02}. Again the crucial derivation of a rate equation 
model from a kinetic equation is missing. In this paper, we provide such a derivation starting 
from a simple Boltzmann-Poisson system that describes one-dimensional (1D) electron 
transport in the lowest miniband of a strongly coupled SL:
\begin{eqnarray} 
&&{\partial f\over \partial t} + v(k) {\partial f\over \partial x} +  {eF\over \hbar} 
{\partial f\over \partial k} = - \nu_{e}\,  \left(f - f^{FD}\right) \nonumber\\
&&\quad - \nu_{i}\, {f(x,k,t) - f(x,-k,t)\over 2}  ,  \label{1}\\
&&\varepsilon\, {\partial F\over\partial x} = {e\over l}\, (n-N_{D}),  \label{2}\\   
&& n = { l\over 2\pi} \int_{-\pi/l}^{\pi/l} f(x,k,t) dk =
{ l\over 2\pi} \int_{-\pi/l}^{\pi/l} f^{FD}(k;n) dk,\quad \label{3}\\
&& f^{FD}(k;n) = {m^{*}k_{B}T\over\pi
\hbar^2}\, \ln\left[1+ \exp\left({\mu - {\cal E}(k)\over k_{B}T}\right)\right] .
\label{4}
\end{eqnarray} 
Here $l$, $\varepsilon$, $f$, $n$, $N_{D}$, $k_{B}$, $T$, $F$, $m^*$ and $e>0$ are the 
SL period, the dielectric constant, the
one-particle distribution function, the 2D electron density, the 2D doping density, the 
Boltzmann constant, the lattice temperature, minus the electric field, the effective mass of 
the electron, and minus the electron charge, respectively. The first collision term represents 
energy relaxation towards a 1D effective Fermi-Dirac distribution $f^{FD}(k;n)$ 
(local equilibrium) \cite{wac02} with collision frequency $\nu_{e}$. The second collision 
term accounts for impurity elastic collisions: $\int_{-\pi/l}^{\pi/l}\phi_{0}(x,k,k') 
\delta({\cal E}(k) - {\cal E}(k')) (f(k')- f(k)) dk' = 2\phi_{0}(x,k,-k) [f(-k) - f(k)]/(
\Delta l \sin kl) \equiv \nu_{i} [f(-k) - f(k)]/2$, provided we use the tight-binding 
miniband dispersion relation, ${\cal E}(k)=(\Delta/2)\, (1-\cos kl)$ ($\Delta$ is the 
miniband width), and ignore transversal degrees of freedom \cite{ger93}. For simplicity, 
$\nu_{e}$ and $\nu_{i}$ will be fixed constants.

\begin{figure}
\begin{center}
\includegraphics[width=7cm,angle=270]{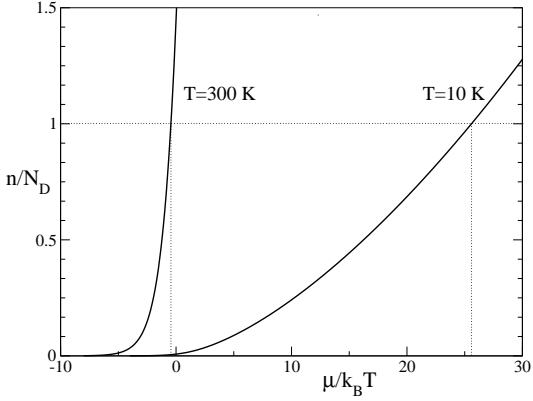}
\vskip -1.5cm
\caption{Electron density versus chemical potential at 10 K ($M =25.59 k_{B}T$) and at 
300 K ($M= -0.45 k_{B}T$). } 
\label{fig1}
\end{center}
\end{figure}

Exact and Fermi-Dirac distribution functions have the same electron density, thereby 
preserving charge continuity as in the BGK (Bhatnagar-Gross-Krook) models of collision 
processes \cite{bha54}. Then the chemical potential $\mu$ depends on $n$ and is found by 
inverting the exact relation (\ref{3}); cf.\ Fig.~\ref{fig1}. BGK collision terms with a 
Boltzmann distribution function, the {\em Boltzmann limit} of Eq.\ (\ref{4}), were 
introduced by Ignatov {\it et al.\ } \cite{ign87}, who adapted collision models by 
Ktitorov {\it et al.\ } containing inelastic energy relaxation and elastic impurity 
momentum relaxation terms \cite{kti72}.

To derive a reduced balance equation for $n$, we shall assume that the electric field 
contribution in Eq.\ (\ref{1}) is comparable to the collision terms and that these terms 
dominate the other two. This is the so-called {\em hyperbolic limit}, in which the ratio of 
$\partial f/\partial t$ or $v(k)\partial f/\partial x$ to $(eF/\hbar)\,\partial f/
\partial k$ is of order $\epsilon\ll 1$. Let $v_{M}$ and $F_{M}$ be electron velocity 
and field scales typical of the macroscopic phenomena described by the sought balance 
equation; for example, let them be the positive values at which the (zeroth order) drift velocity 
reaches its maximum. In the hyperbolic limit, the time $t_{0}$ it takes an electron with speed 
$v_{M}$ to traverse a distance $x_{0}=\varepsilon F_{M}l/(eN_{D})$, over which the 
field variation is of order $F_{M}$, is much longer than the mean free time between 
collisions, $\nu_{e}^{-1}\sim \hbar/(eF_{M}l)=t_{1}$. We therefore define $\epsilon
=t_{1}/t_{0}=\hbar v_{M}N_{D}/(\varepsilon F_{M}^2 l^2)$ and formally multiply
the two first terms on the left side of (\ref{1}) by $\epsilon$. \cite{epsilon} After
obtaining the number of desired terms, we set $\epsilon =1$. The solution of Eq.\ 
(\ref{1}) for $\epsilon =0$ is straightforwardly calculated in terms of its Fourier 
coefficients as $f^{(0)}(k;n) = \sum_{j=-\infty}^{\infty} f^{(0)}_{j} e^{ijkl}$, with 
$f^{(0)}_{j} = (1-ij {\cal F}/\tau_{e})\,f^{FD}_{j}/(1 + j^2 {\cal F}^{2})$, in which 
${\cal F} = F/F_{M}$, $F_M = \hbar\sqrt{\nu_{e}(\nu_{e}+\nu_{i})}/(el)$,  and 
$\tau_{e}= \sqrt{(\nu_{e}+\nu_{i})/\nu_{e}}$. Since $f^{FD}$ is an even function of 
$k$, its Fourier coefficient $f^{FD}_{j}$ is real. Note that Eq.\ (\ref{3}) implies 
$f^{(0)}_{0} = f^{FD}_{0} = n$. 

We shall derive a reduced balance equation for the electron density by using the 
Chapman-Enskog ansatz \cite{bon00}:
\begin{eqnarray} 
&& f(x,k,t;\epsilon) = f^{(0)}(k;n) + \sum_{m=1}^{\infty} f^{(m)}(k;n)\, 
\epsilon^{m} ,    \label{12}\\
&&  {\partial n\over\partial t} = \sum_{m=0}^{\infty}  N^{(m)}(n)\, 
\epsilon^{m}.  \label{13}
\end{eqnarray} 
The coefficients $f^{(m)}(k;n)$ depend on the `slow variables' $x$ and 
$t$ only through their dependence on the electron density and the electric field (which is itself 
a functional of $n$). The electron density obeys a reduced evolution equation (\ref{13})
in which the functionals $N^{(m)}(n)$ are chosen so that the $f^{(m)}(k;n)$ are bounded 
and $2\pi/l$-periodic in $k$. Moreover the condition,
$\int_{-\pi/l}^{\pi/l} f^{(m)}(k;n) \, dk = 2\pi\, f^{(m)}_{0}/l= 0,  \quad m\geq 1$, 
ensures that $f^{(m)}$, $m\geq 1$, do not contain contributions proportional to the 
zero-order term $f^{(0)}$. $N^{(m)}(n)$ can be found by integrating (\ref{1}) over $k$,
using (\ref{3}), and inserting (\ref{12}) in the result: $N^{(m)}(n)=- l\, (\partial/
\partial x)\int_{-\pi/l}^{\pi/l} v(k) f^{(m)}dk/(2\pi)$. Then, integration of (\ref{2})
over $x$ yields
\begin{eqnarray} 
\varepsilon {\partial F\over\partial t} + {e\over 2\pi}\,\sum_{m=0}^\infty
\epsilon^m\, \int_{-\pi/l}^{\pi/l} v(k)\, f^{(m)}(k;n) \, dk = J(t),
\label{14}  
\end{eqnarray} 
where $J(t)$ is the total current density. To find the equations for $f^{(m)}$, we insert 
(\ref{12}) and (\ref{13}) in (\ref{1}), and equate like powers of $\epsilon$:
\begin{eqnarray} 
{\cal L} f^{(1)} &=&  \left.  -
 \left({\partial \over \partial t} + v(k) {\partial \over \partial x}\right) 
 f^{(0)}\right|_{0} , \label{15}\\
 {\cal L} f^{(2)} &=&  \left.  \left. -  \left({\partial \over \partial t} 
 + v(k) {\partial \over \partial x}\right) f^{(1)}\right|_{0} - {\partial 
 \over \partial t} f^{(0)}\right|_{1}, \quad \label{16}
\end{eqnarray} 
and so on. We have defined ${\cal L} u(k) \equiv eF\hbar^{-1} du(k)/dk + ( \nu_{e} + 
\nu_{i}/2) u(k) + \nu_{i} u(-k)/2$, and the subscripts 0 and 1 mean that $\partial n/
\partial t$  is replaced by $N^{(0)}(n)$ and by $N^{(1)}(n)$, respectively. 

 The linear equation ${\cal L} u= S$  has a bounded $2\pi/l$-periodic solution provided 
 $\int_{-\pi/l}^{\pi/l}S\, dk =0$. This solvability condition together with Eqs.\ 
 (\ref{15}), (\ref{16}), etc.\ also yield the previously found $N^{(m)}$ and the 
 reduced equation (\ref{14}). Keeping only the leading order terms in (\ref{14}), we
 obtain
\begin{eqnarray} 
&& \varepsilon {\partial F\over\partial t} + e\, n\,Ê{\cal M}\left({n\over
N_{D}}\right)\, v_{M}\, V({\cal F})= J(t), \label{6}\\ 
&& V({\cal F}) = {2{\cal F}\over 1 + {\cal F}^2} , \quad
v_{M} = {\Delta l\, {\cal I}_{1}(M)\over 4\hbar\tau_{e} {\cal I}_{0}(M)},  
\label{10}\\
&& {\cal I}_{m}(s) = \int_{-\pi}^{\pi}\cos (m k)\,\ln\left( 1+e^{s-\delta+
\delta\cos k}\right)\,dk, \label{8}
\end{eqnarray} 
provided ${\cal M}(n/N_{D}) = {\cal I}_{1}(\tilde{\mu})\, {\cal I}_{0}(M)/[ 
{\cal I}_{1}(M) {\cal I}_{0}(\tilde{\mu})]$, $\tilde{\mu}\equiv\mu/(k_{B}T)$, 
and $\delta=\Delta/(2 k_{B}T)$. Using Eq.\ (\ref{3}), the dimensionless chemical 
potential $\tilde{\mu}= \tilde{\mu}(n/N_{D})$ is calculated graphically in 
Fig.~\ref{fig1} as a function of $n/N_{D}$, with $\tilde{\mu}(1)=M$. Then we have 
${\cal M}(1)  =1$. In the Boltzmann limit, ${\cal M}=1$ for any $n$, and the electron 
current density in (\ref{6}) has the usual drift form. Thus ${\cal M}$ is a low-temperature, 
density-dependent correction to the usual drift current density. The drift velocity, $v_{M} 
V({\cal F})$, has the Esaki-Tsu form with a maximum that becomes $v_{M}\approx 
\Delta l I_{1}(\delta)\sqrt{\nu_{e}}/[4\hbar I_{0}(\delta)\sqrt{\nu_{e}+\nu_{i}}
]$ in the Boltzmann limit \cite{ign87} ($I_{n}(\delta)$ is the modified Bessel function 
of the $n$th order).

The first-order correction in (\ref{14}) is found by first solving (\ref{15}). After 
straightforward but lengthy calculations and setting $\epsilon=1$, we obtain (here $g'$ 
means $dg/dn$):
\begin{eqnarray}
&&\varepsilon {\partial F\over\partial t} + {\cal V}\left(F,{\partial F\over
\partial x}\right)\, {eN_{D}\over l}\, \left( 1 + {\varepsilon l\over eN_{D}}\,
{\partial F\over\partial x}\right)\nonumber\\
&& \quad =  D\left(F,{\partial F\over\partial x}\right)\,
\varepsilon {\partial^2 F\over \partial x^2}
 + A\left(F,{\partial F\over\partial x}\right)\, J(t) ,   \label{22}\\
&& A = 1 + {2 e v_{M} F_{M}^3\, [F_{M}^2 - (1+2 \tau^{2}_{e})\, F^2]\over 
\varepsilon l (\nu_{e}+ \nu_{i}) (F_{M}^2 + F^{2})^3 }\, n {\cal M},  \label{24}\\
&& {\cal V} = v_{M}V {\cal M} \left( A - {\Delta B\over  2 e  }\, 
{\partial F\over\partial x} \right) ,   \label{25}\\
&& D = {\Delta^2 l F_{M}\over 8\hbar e\tau_{e} \, (F^2_{M} + F^{2}) }  
 \left( 1 - {4\hbar v_{M}C\over\Delta l} \right) ,  \label{26}\\
&& B = { (5F^2_{M}-4F^2) {\cal M}_{2}\over (F^2_{M}+ 4F^2)^2 {\cal M}} 
\nonumber\\
&& \quad - {4\hbar v_{M} F^2_{M}(F^2_{M}-F^2) (\tau_{e} + \tau^{-1}_{e}) 
(n{\cal M})' \over \Delta l (F^2_{M}+F^2)^3}, \label{27}\\
&& C = {\tau_{e}  (F^2_{M}-2F^2)  (n {\cal M}_{2})' \over F^2_{M} +  4 F^2 }
+ {8\hbar v_{M} [F F_{M}(n {\cal M})']^2\over \Delta l ( F^2_{M}+F^2)^2}
.  \quad \quad   \label{9}
\end{eqnarray} 
Here the density-dependent function ${\cal M}_{2}(n/N_{D})= {\cal I}_{2}(\tilde{
\mu})\, {\cal I}_{0}(M)/[{\cal I}_{0}(\tilde{\mu})\, {\cal I}_{1}(M)]$ becomes 
simply the constant $I_{2}(\delta)/I_{1}(\delta)$ in the Boltzmann limit. Despite its 
formidable appearance, 
the {\em generalized drift-diffusion equation} (GDDE) (\ref{22}) is (in dimensionless 
units) a small perturbation of the drift equation (\ref{6}), analyzed in studies of the Gunn 
effect a long time ago \cite{kni66,epsilon}. Table \ref{table1} shows that the solution of 
the GDDE and (\ref{2}) yield self-oscillations of the current with frequencies that agree 
with those measured by Schomburg {\it et al} \cite{sch98}. 

\begin{table}[ht]
\caption{Numerical values of the oscillation frequencies
$\nu_{\rm num}$, compared with the experimental value
$\nu_{\rm exp}$ for five of the SLs of Ref.~\onlinecite{sch98},
together with the corresponding applied voltage $\Phi$. $d_{W}$ and $d_{B}$
are well and barrier widths, respectively.}
\begin{center} \footnotesize
\begin{tabular}{|c|c|c|c|c|c|}
 \hline
$d_{W}$ (\AA) & $d_{B}$ (\AA) & $N_{D}/l$ (cm$^{-3}$) & $\nu_{\rm exp}$ (GHz) &
$\nu_{\rm num}$ (GHz) & $\Phi$ (V)\\
 \hline
 \hline
 51.3 & 8.7 & 1.4 $\times 10^{17}$& 19.44 & 19.5 & 0.95 \\
48 & 9 & 8 $\times 10^{16}$& 29.12 & 29.1 & 1.07 \\
40 & 10 &  8 $\times 10^{16}$ & 46.35 & 46.5 & 1.2 \\
36.4 & 9.3  & $10^{17}$ & 52.79 & 52.8 & 1.24 \\
 35.4 & 9.6 &  9 $\times 10^{16}$ & 65 &  65  & 1.73\\
 \hline
\end{tabular}
\end{center}
\label{table1}
\end{table}

\begin{figure}
\begin{center}
\includegraphics[width=8cm,angle=0]{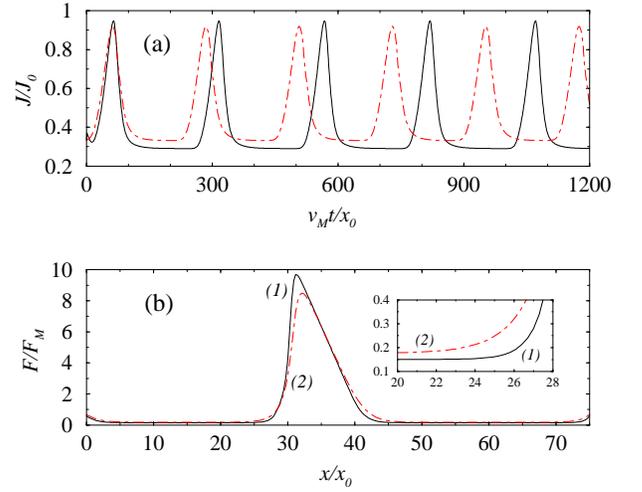}
\caption{(a) Current ($J_{0}=ev_{M}N_{D}/l$) vs.\ time during self-oscillations for a 
100-period SL at 300K, as described by the GDDE in the Boltzmann limit (solid 
line) and by the DDE (dashed line). (b) Comparison between the 
dipole wave for the DDE, (1), and the dipole wave for the GDDE, (2). }
\label{fig4}
\end{center}
\end{figure}

An often used DDE consists of Eq.\ (\ref{6}) (with ${\cal M}\equiv 1$) plus a diffusion 
term obeying the Einstein relation \cite{sch02}:
\begin{eqnarray} 
\varepsilon\, {\partial F\over \partial t} + {e v_{M} n\over l}\, 
V({\cal F}) = J(t) + {k_{B}T v_{M}\over Fl}\, V({\cal F}) {\partial n\over 
\partial x}\,.  \label{31}
\end{eqnarray} 
The difference between the predictions of (\ref{22}) and (\ref{31}) can be remarkable if 
the dimensionless parameter $\epsilon$ is relatively large and the dimensionless coefficient 
$\delta=\Delta/(2k_{B}T)$ is not small, as illustrated in Fig.~\ref{fig4}. The parameter 
values in this figure correspond to the 5.13-nm GaAs/0.87-nm AlAs SL of 
Ref.~\onlinecite{sch98}, for which $\epsilon=0.34$: $N_{D} = 0.84 \times 
10^{11}$cm$^{-2}$, $\Delta=43$meV, $\nu_{e}=\nu_{i}= 10^{13}$Hz, $x_{0}/l = 
0.75$. We have selected bias and boundary conditions so that dipole mediated current 
self-oscillations occur in this SL: voltage bias divided by SL length equals 1.2 $F_{M}$, 
and ${\cal F}= 2 J l/(eN_{D}v_{M})$ at both SL ends. The difference in oscillation 
frequency and wave shape can be explained by taking into account the equal-area rule as 
in the theory of the Gunn effect \cite{bon75}: the taller wave of the GDDE moves at a 
slower average speed than the wave of (\ref{31}). 

\begin{figure}
\begin{center}
\includegraphics[width=7cm,angle=270]{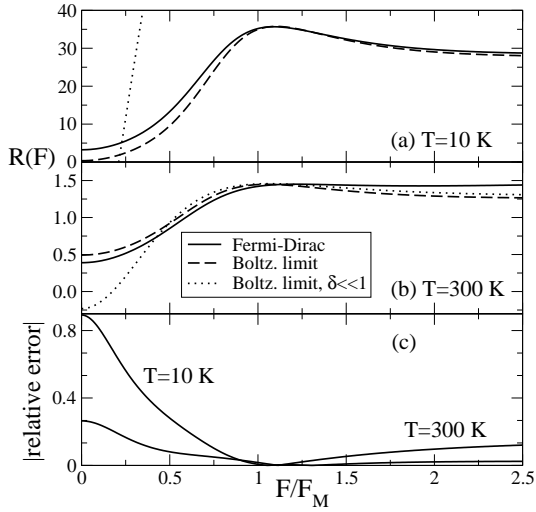}
\vskip -1cm
\caption{Ratio $R(F)$ at: (a) 10 K, and (b) 300 K. 
(c) Relative error of the Boltzmann limit result with respect to using the Fermi-Dirac 
distribution. } 
\label{fig3}
\end{center}
\end{figure}

For SLs with a smaller value of $\epsilon$, the difference between the predictions of the 
GDDE and the DDE (\ref{31}) is smaller. Is there a limit in which these equations agree?
To explore this, we calculate the deviation of drift velocity and diffusion coefficient in the 
GDDE from the Einstein relation (setting $n=N_{D}$ and $\tilde{\mu} = M$):
\begin{eqnarray} 
 && R(F)\equiv {eF\, D(F,0)\over k_{B}T\, {\cal V}(F,0)} = { \Delta {\cal I}_{0}
 \over 4 k_{B}T {\cal I}_{1}}\nonumber\\
 && \times {1- {1-2{\cal F}^2 \over 1+ 4{\cal F}^2} \left({n{\cal I}_{2}
 \over {\cal I}_{0} }\right)'  -  {2{\cal F}^2 \over (1+ 
 {\cal F}^2)^2\tau_{e}^2} \left[\left({n{\cal I}_{1}\over {\cal I}_{0} }
 \right)'\right]^2 \over 1+ {e^2\Delta l N_{D} [1- (1+\tau_{e}^2) 
{\cal F}^2] {\cal I}_{1}\over 2\varepsilon \hbar^{2} (\nu_{e}+\nu_{i})^2 
{\cal I}_{0}\, (1+{\cal F}^2)^3} } .    \label{29}
\end{eqnarray} 
In the Boltzmann limit, exp$(\tilde{\mu} - \delta + \delta\cos kl) \ll 1$, and we can 
substitute the modified Bessel functions $I_{s}(\delta)$ instead of ${\cal I}_{s}(M)$ in 
the previous formula. Moreover, Eqs.\ (\ref{3}) and (\ref{4}) give $n\approx
 e^{\tilde{\mu}-\delta}\, I_{0}(\delta)\, \rho_{0}N_{D}$, where $\rho_{0}=
m^* k_{B}T/(\pi\hbar^2N_{D})$. For $n=N_{D}$, the Boltzmann limit holds provided 
$\rho_{0}\gg 1$. If we also have $\delta=\Delta/(2k_{B}T)\ll 1$, (\ref{29}) becomes 
\begin{eqnarray} 
&& R(F) \sim 1 + {\Delta^2\over 8 k^2_{B} T^2}\left[ {3F^2\over 2(F^2_{M}+
4F^2)} - {F_{M}^2/\tau_{e}^2\over (F_{M}^2 +F^2)^2} \right. \quad\,
\nonumber\\
&& \left. \times \left({k_{B}TN_{D} [F_{M}^2-(1+\tau_{e}^2)F^2]\over \varepsilon l
\, (F_{M}^2+F^2)} + F^2\right) \right].  \label{30}
\end{eqnarray} 
Ignoring correcting terms, in this limit the right hand side of Eq.\ (\ref{30}) becomes 1 
and the Einstein relation holds. Fig. \ref{fig3} shows the deviation from the Einstein 
relation at different temperatures, either using the Fermi-Dirac distribution in Eq.\ 
(\ref{1}), using its Boltzmann limit, or the two-term approximation (\ref{30}). 
Deviations are more appreciable at low temperatures. In the limit $k_{B}T\gg $max
$(\Delta,\pi\hbar^2N_{D}/m^*)$, the GDDE (\ref{22}) becomes the DDE (\ref{31}) 
up to terms of order $\epsilon\delta$ if we set $A=1+O(\epsilon)\sim 1$.

In conclusion, we have derived a generalized drift-diffusion model for charge transport in
miniband superlattices by means of a consistent Chapman-Enskog method. At all 
temperatures, its predictions deviate appreciably from those 
of the usual DDE with the Esaki-Tsu drift velocity and diffusion obeying the 
Einstein relation. The DDE holds in the limit  $\epsilon\ll 1$, $k_{B}T\gg $max
$(\Delta,\pi\hbar^2N_{D}/m^*)$, which is not very realistic for many strongly coupled SLs, even at 
room temperature. Detailed analyses and comparison between the predictions of the two DD 
models and those of the original kinetic equation will be considered in future works.

%\acknowledgments
This work has been supported by the MCyT grant BFM2002-04127-C02-01, and by the 
European Union under grant HPRN-CT-2002-00282.

\end{document}